\documentclass[aps,a4paper,10pt,twocolumn]{revtex4} 
\usepackage{amsfonts}
\usepackage{amssymb}
\usepackage{mathrsfs}
\usepackage[mathscr]{euscript}
\usepackage[dvips]{graphicx}
\usepackage{fancyhdr}
\usepackage[hypertex=true]{hyperref}
\usepackage{makeidx}

\begin{document}

\title{Theory of directional pulse propagation}
\author{P. Kinsler}
\affiliation{
  Department of Physics, Imperial College London,
  Prince Consort Road,
  London SW7 2BW, 
  United Kingdom.}
\author{S.B.P. Radnor}
\affiliation{
  Department of Physics, Imperial College London,
  Prince Consort Road,
  London SW7 2BW, 
  United Kingdom.}
\author{G.H.C. New}
\affiliation{
  Department of Physics, Imperial College London,
  Prince Consort Road,
  London SW7 2BW, 
  United Kingdom.}

\begin{abstract}

We construct combined electric and magnetic field variables
 which independently represent energy flows in
 the forward and backward directions respectively, 
 and use these to re-formulate Maxwell's equations.
These variables enable us to not only judge the effect
 and significance of backward-travelling field components,
 but also to discard them when appropriate.
They thereby have the potential to simplify numerical simulations, 
 leading to potential speed gains of up to 100\% 
 over standard FDTD or PSSD simulations.
We present results for various illustrative situations,
 including an example application to second harmonic generation in 
 periodically poled lithium niobate.
These field variables are also used to derive both
 envelope equations useful for narrow-band pulse propagation,
 and a second order wave equation.
Alternative definitions are also presented.

\end{abstract}

\pacs{X}



\lhead{\includegraphics[height=5mm,angle=0]{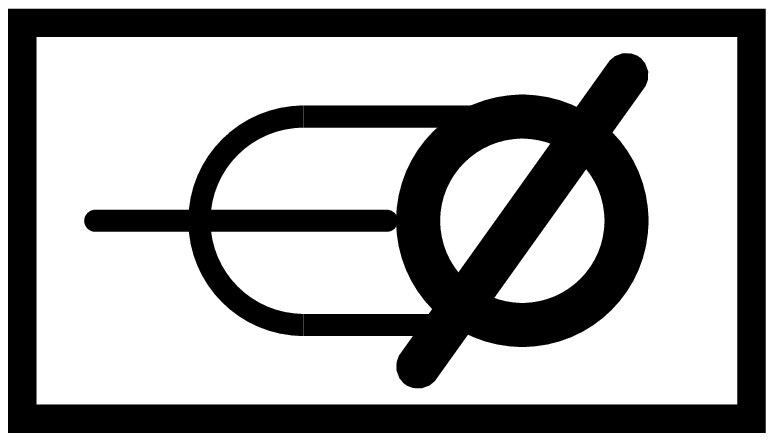}~~FLVAR}
\chead{~}
\rhead{
\href{mailto:Dr.Paul.Kinsler@physics.org}{Dr.Paul.Kinsler@physics.org}\\
\href{http://www.kinsler.org/physics/}{http://www.kinsler.org/physics/}
}

\date{\today}
\maketitle
\thispagestyle{fancy}



\chead{Directional pulse propagation}


{\em
Published as
Phys. Rev. A{\bf 72}, 063807 (2005).
A more detailed derivation of the fields and equations 
herein can be found at http://arxiv.org/abs/physics/0611216
}

\section{Introduction}\label{S-introduction}

We introduce electro-magnetic field variables $\vec{G}^{\pm}$ 
 that are designed to have directional characteristics. 
These variables have the potential to
 speed up numerical simulations, 
 while providing valuable insight into the
 process of optical pulse propagation at the same time.     
Simple plane-polarized versions of $\vec{G}^{\pm}$ for a dispersionless medium 
 were originally proposed by 
 Fleck at the beginning of ref. \cite{Fleck-1970prb}, 
  although he did not use them in the rest of the paper.
In the generalized form defined below, it is possible to use them
   to advantage in practical situations.
We note that a different approach to directional pulse propagation 
   based on projection operators was proposed by
   Kolesik et.al. \cite{Kolesik-MM-2002prl,Kolesik-M-2004pre}; 
   there is also the recent work of 
   Ferrando et.al. \cite{Ferrando-ZCBM-2005pre} based on a second order 
   wave equation.

The essential characteristic of $\vec{G}^{+}$ and $\vec{G}^{-}$
 is that they represent energy fluxes directed 
 in the forward and backward directions respectively.
This implies that $\vec{G}^{+}$ is the appropriate variable to use 
  in situations where pulses are travelling only in the forward direction.
Indeed, 
  as we will explain, 
  optimal construction of $\vec{G}^{+}$ 
  makes $\vec{G}^{-}$ negligible under these circumstances, 
  and the computational effort can then be halved by
  neglecting $\vec{G}^{-}$ altogether.

If we apply a $z$-propagated 
 pseudospectral spatial-domain (PSSD)
 algorithm \cite{Tyrrell-KN-2005jmo}, 
 we also gain a 
 fast and flexible treatment of dispersion and nonlinear effects, 
  which significantly outperforms standard
 finite difference time-domain (FDTD)
 methods \cite{Yee-1966iee,Joseph-T-1997ieee}. 
Further, 
  since many authors (including the recent \cite{Kolesik-M-2004pre})
 assume that the backward field is negligible in any case, 
 the explicit appearance of $\vec{G}^{-}$ within our
 formalism provides a direct test of the validity of this assumption.
A further advantage of $\vec{G}^{\pm}$ is that it is as easy
 to include magneto-optic effects as  
 electro-optic effects such as dispersion and nonlinearity.
It is in situations involving both electro- and magneto-optic 
 effects where we achieve the greatest computational speed increase -- 
 potentially up to 100\% faster. 
Moreover, 
  even if one chooses to propagate an optical pulse using $E$ and $H$,
  it is still easy to analyse its directional characteristic by
  constructing $\vec{G}^{\pm}$ after the event.

After reviewing Fleck's original form of the $\vec{G}^{\pm}$ variables 
 at the start of section \ref{S-definitions}, 
  we proceed to discuss how to represent the
  permittivity and permeability of the medium; 
this is a crucial step in the optimal construction of $\vec{G}^{\pm}$ 
  in a generalized form.
The treatment of nonlinearities and the calculation of energy and flux are
  also covered.

In Section \ref{S-firstorder}, we derive a first-order wave equation,
  both in a form that is fully equivalent to Maxwell's equations, 
  and in a more useful one that is applicable in the transverse field limit.
In section \ref{S-simulations}, we demonstrate a simple procedure
  for the numerical implementation of $\vec{G}^{\pm}$ simulations;
  techniques for specifying initial conditions and for handling
  dispersion and nonlinear effects are examined in detail.
We take as an example the case of 
  second harmonic generation in periodically-poled lithium niobate, 
  to demonstrate that our method can be applied to practical 
  as well as illustrative simulations.
In all cases, 
  we retain both $\vec{G}^{+}$ and $\vec{G}^{-}$,
 but show that with optimal construction, 
 $\vec{G}^{-}$ can be made negligible.

In section \ref{S-envelopes} we derive a propagation equation for 
 envelopes based on $\vec{G}^{\pm}$; 
 in section \ref{S-secondorder} we develop a second order wave equation; 
 and in section \ref{S-alternatives} we propose alternative definitions
 for field variables with directional properties.
Finally, in section \ref{S-conclusion}, we present our conclusions.

\section{Definitions}\label{S-definitions}

For plane-polarized fields, 
   propagating in the $z$ direction in a dispersionless medium, 
   Fleck defined the direction field variables
~
\begin{eqnarray}
  {G}^{\pm}
&=&
  \sqrt{\epsilon} E_x
 \pm 
  \sqrt{\mu} H_y
.
\label{eqn-S-defs-GFleck}
\end{eqnarray}
Their directional properties are apparent from the 
 form of the Poynting vector
~
\begin{eqnarray}
  S = E_x H_y
&=&
  \frac{1}{4\sqrt{\epsilon\mu}}
  \left[
    {G}^{+2} - {G}^{-2}
 \right]
,
\label{eqn-S-defs-SFleck}
\end{eqnarray}
which shows that ${G}^{+}$ and ${G}^{-}$ are associated with 
 positive and negative energy flux respectively.
Unfortunately, 
  if eqn. (\ref{eqn-S-defs-GFleck}) is used to describe
  a forward-propagating pulse in a {\em dispersive} medium,
  it will contain significant contributions from 
  both ${G}^{+}$ and ${G}^{-}$.
We therefore need to generalize the construction in order to
  make the concept useful in practical situations.

\subsection{Medium Parameters}\label{Ss-definitions-ref}

The definitions of ${G}^{\pm}$ 
 (and their generalized vector counterparts $\vec{G}^{\pm}$, introduced below)
 depend on the properties of
 the propagation medium through the
 permittivity $\epsilon$ and permeability $\mu$.  
In principle it would be attractive to define $\vec{G}^{\pm}$ using
 the exact values of $\epsilon, \mu$ (including the nonlinearity), 
 but this is usually impractical, 
 and we will instead use ``reference'' values $\epsilon_r$, $\mu_r$, 
 chosen to be as close as practicable
 to the true medium properties,
 typically by including all the dispersive properties.

In the frequency domain (indicated by tildes), we write
~
\begin{eqnarray}
  \tilde{\epsilon} 
~~~~ = \tilde{\epsilon}_r(\omega) + \tilde{\epsilon}_c(\omega)
&=&
   \tilde{\alpha}_r^2(\omega) 
 + \tilde{\alpha}_r(\omega) ~ \tilde{\alpha}_c(\omega),
\label{eqn-defs-alphaX}
\\
  \tilde{\mu} 
~~~~ = \tilde{\mu}_r(\omega)  + \tilde{\mu}_c(\omega)
& =&
    \tilde{\beta}_r^2(\omega) 
  + \tilde{\beta}_r(\omega) ~ \tilde{\beta}_c(\omega),
\label{eqn-defs-betaX}
\end{eqnarray}
where the correction parameters $\tilde{\epsilon}_c$ and $\tilde{\mu}_c$
 represent the discrepancy between the true values and the reference.
The smaller these correction terms are, 
 the better the match, 
 and the more likely it is that a description involving only ${G}^{+}$ 
 will suffice.
Note that since the definitions of ${G}^{\pm}$ depend on the square roots of 
 $\tilde{\epsilon}$ and $\tilde{\mu}$, 
 we introduce the $\tilde{\alpha}$ and $\tilde{\beta}$ parameters, 
 which will feature prominently 
 (along with their time domain counterparts $\alpha, \beta$),
 in the generalized definitions of 
 $\vec{G}^{\pm} $ that follow.

By using these frequency dependent parameters in the generalized
  definitions of ${G}^{\pm}$, 
  we are able to propagate pulses using only the ${G}^{+}$ variable,
  a gain in both mathematical simplicity and computational speed.

\subsection{$\vec{G}^{\pm} $ variables }\label{Ss-definitions-G}

The generalized definitions of the $\vec{G}^{\pm}$ variables 
 in the frequency and time domains are
~
\begin{eqnarray}
  \vec{G}^{\pm} (\omega)
&=&
  \tilde{\alpha}_r (\omega)  \vec{E}  (\omega)
 \pm 
  \vec{u} \times \tilde{\beta}_r (\omega)  \vec{H} (\omega)
,
\label{eqn-S-defs-GvectorW}
\\
{G}^{\circ}  (\omega)
&=&
  \vec{u} 
  \cdot
  \left[
    \tilde{\beta}_r (\omega)  \vec{H}  (\omega)
  \right]
;
\label{eqn-S-defs-GcircW}
\\\
\textrm{or} ~~~~ ~~~~
  \vec{G}^{\pm} (t)
&=&
  \alpha_r (t)  \ast \vec{E}  (t)
 \pm 
  \vec{u} \times \beta_r (t) \ast \vec{H} (t)
,
\label{eqn-S-defs-Gvector}
\\
{G}^{\circ}  (t)
&=&
  \vec{u} 
  \cdot
  \left[
    \tilde{\beta}_r (t) \ast  \vec{H}  (t)
  \right]
,
\label{eqn-S-defs-Gcirc}
\end{eqnarray}
where $\vec{u}$ is the unit vector in the direction of propagation, 
 and $\alpha_{r}(t)$ and $\beta_{r}(t)$ are the (inverse) Fourier transformed
 versions of $\tilde{\alpha}_r (\omega)$ and $\tilde{\beta}_r (\omega)$.
The symbol ``$\ast$'' is used to denote a convolution: 
 $a \ast b  = \int a(t-t') b(t) dt'$.
The variable ${G}^{\circ}$ involves 
 the longitudinal part of the magnetic field, 
 which is eliminated in the $\vec{u} \times \vec{H}$ operation in 
 eqn. (\ref{eqn-S-defs-GvectorW}).
Although we will generally make a transverse approximation
 in which ${G}^{\circ}=0$ and $\vec{u} \cdot \vec{G}^{\pm} =0$, 
 we retain the longitudinal parts of the field to ensure a 
 complete description.
To avoid cluttering the notation, 
 we do not apply tildes to the spectral forms of the field quantities
 $\vec{G}^{\pm}$, ${G}^{\circ}$, $\vec{E}$, $\vec{H}$, 
 and rely instead on the arguments ($t$ or $\omega$) or the context, 
 to distinguish between domains.

Inverting eqn. (\ref{eqn-S-defs-GvectorW}) gives the following 
 expressions for the electric and magnetic fields as a function of 
 $\vec{G}^{\pm}$ and ${G}^{\circ}$
~
\begin{eqnarray}
  \vec{E} (\omega)
&=&
  \frac{1}{2 \tilde{\alpha}_r (\omega)} 
  \left[ \vec{G}^{+} (\omega) + \vec{G}^{-} (\omega) \right]
,
\label{eqn-S-defs-Evector}
\\
  \vec{H}  (\omega)
&=& 
  \frac{1}{2 \tilde{\beta}_r (\omega)} 
  \vec{u} 
  \times
  \left[ 
    \vec{G}^{+} (\omega) - \vec{G}^{-} (\omega) 
  \right]
 +
  \frac{
  \vec{u}
  {G}^{\circ} (\omega) 
       }
       {\tilde{\beta}_r (\omega)} 
.
~~~~
\label{eqn-S-defs-Hvector}
\end{eqnarray}

The divergence of $\vec{G}^{\pm}$, 
 allowing for both charge density $\rho$ and 
 current density $\vec{J}$, is
\begin{eqnarray}
  \nabla \cdot \vec{G}^{\pm}
&=&
  \frac{\tilde{\alpha}_r}
       {\tilde{\alpha}^2}
  \rho
 \pm
  \frac{\imath \omega}{2}
  \frac{\tilde{\alpha}^2}
       {\tilde{\alpha}_r}
  \tilde{\beta}_r 
  \vec{u} \cdot 
  \left(
    \vec{G}^{+}
   +
    \vec{G}^{-}
  \right)
 \mp
  \tilde{\beta}_r 
  \vec{u} \cdot \vec{J}
. ~~~
\label{eqn-S-defs-divergenceGpm}
\end{eqnarray}
We note that this is zero when the $\rho$ and $\vec{J}$ are zero,
 as long as there is no longitudinal electric field. 

Different choices of $\tilde{\epsilon}_r, \tilde{\mu}_r$ 
 produce different $\vec{G}^{\pm}$ pairs.
Whilst using the true values to describe a forward propagating pulse 
 results in $\vec{G}^{-}=0$,
 any other choice of reference will produce a non-zero $\vec{G}^{-}$ component
 that co-propagates with $\vec{G}^{+}$.
Note that this $\vec{G}^{-}$ still has an energy flux directed in the 
 reverse direction ($-\vec{u}$),
 but travels forwards with the $\vec{G}^{+}$ with which it is tightly coupled.

We will almost always choose
 $\tilde{\epsilon}_r, \tilde{\mu}_r$ to include the entire
 linear dispersion of the medium.
We exclude the nonlinearity because it removes the ability to
  reconstruct $\vec{E}, \vec{H}$ fields uniquely from the $\vec{G}^{\pm}$,
 as can be seen from 
 eqns. (\ref{eqn-S-defs-GvectorW}, \ref{eqn-S-defs-Gvector}, 
        \ref{eqn-S-defs-Evector}, \ref{eqn-S-defs-Hvector}), 
 which will become nonlinear in $\vec{E}$ and $\vec{H}$.

The vectorized definitions of $\vec{G}^{\pm}$ 
 accommodate any polarization of the $E$ and $H$ fields. 
For propagation along the $z$ axis, the $x$ component of 
 $\vec{G}^{\pm}$ ($G_x^{\pm}$) will contain $E_x$ and $H_y$; 
 and similarly $G_y^{\pm}$, will contain $E_y$ and $H_x$.
It is then a simple matter to see how linearly or circularly polarized
 $\vec{E}$ and $\vec{H}$ fields can be represented 
 in terms of $\vec{G}^{\pm}$.
The definitions are also easily generalized to include birefringent media,
 provided the propagation direction and transverse coordinate axes 
 are such that $\epsilon_r$ and $\mu_r$ become diagonal matrices.

Finally, note that
 $\vec{G}^{\pm}$ bear some resemblance to Beltrami variables 
 (see e.g. \cite{Lakhtakia-1994ijimw,Hillion-1995jpa,Moses-1971siamjam})
 which are defined as
 $\vec{Q} = \sqrt{\epsilon} \vec{E} + \imath \sqrt{\mu} \vec{H}$; 
 but they differ in two important respects.  
First, a given Beltrami $\vec{Q}$ defines $\vec{E}$ and $\vec{H}$ uniquely, 
 whereas both $\vec{G}^{+}$ and $\vec{G}^{-}$ are needed to do the same.
Secondly, $\vec{Q}$ does not assume any preferred 
 direction, whereas the $\vec{G}^{\pm}$ variables include the
 direction $\vec{u}$ in their definition.  
Further, 
 Beltrami variables are not defined using 
 the full time (or frequency) dependence of $\epsilon, \mu$ 
 as we use for $\vec{G}^{\pm}$ in 
 eqns. (\ref{eqn-S-defs-GvectorW}, \ref{eqn-S-defs-Gvector}), 
 although presumably this would be possible.

\subsection{Nonlinearities}\label{Ss-definitions-nonlinear}

Since it is usually impractical to include nonlinearities in the 
 reference parameters,  
 these will normally appear in the correction terms $\epsilon_c$, $\mu_c$.
As an example, 
 consider a $n$-th order (electric) nonlinearity,
 in which case 
 $  \epsilon_c(t) =  \chi^{(n)}(t)
 \ast
  E(t)^{n-1}
$, and 
~
\begin{eqnarray}
  \tilde{\alpha}_c (\omega)
&=&
  \left[ 
    \tilde{\alpha}_r (\omega)
 \right]^{-1}
 .
  \mathscr{F} 
    \left[
      \chi^{(n)}(t) 
     \ast
       E(t)^{n-1}
    \right]
,
\label{eqn-nonlinear-alphac-w}
\\
  \alpha_c (t)
&=&
 \mathscr{F}^{-1}
 \left\{
   \left[ 
     \tilde{\alpha}_r (\omega)
   \right]^{-1}
  .
  \tilde{\chi}^{(n)}(\omega) 
  .
      \mathscr{F}
        \left[    
          E(t)^{n-1}
        \right]
 \right\}
,
\label{eqn-nonlinear-alphac-t}
\end{eqnarray}
where $\mathscr{F}[...]$ is the Fourier transform (FT) from time to frequency,
 and $E(t)$ can be found from eqn. (\ref{eqn-S-defs-Evector}).
If the reference parameters $\tilde{\alpha}_r$ contain
 dispersion (which will be the typical case), 
 we can see from eqn. (\ref{eqn-nonlinear-alphac-w}) that this will make
  $\tilde{\alpha}_c (\omega)$ dispersive even if $\chi^{(n)}$ is 
 instantaneous.
In the case of an instantaneous nonlinearity, 
 this adds more computational work (an extra two FTs), 
 although for non-instantaneous ones we needed the FTs anyway.
If the nonlinearity is instantaneous {\em and} the 
 reference parameters are non-dispersive,
 we have simply 
  $  \alpha_c^{NL}(t) 
  =  \alpha_r^{-n} . \chi^{(n)} . 
     2^{-n+1} \left[ {G}^{+} + {G}^{-} \right]^{n-1}$.

\subsection{Energy and Flux}\label{Ss-definitions-flux}

The $\vec{G}^{\pm}$  are intrinsically directional and do not rely on 
 a carrier wave to impart their directionality.
This becomes clear when the Poynting vector is expressed 
 in terms of $\vec{G}^{\pm }$. 
For transverse fields and dispersive reference parameters, 
 we obtain 
~
\begin{eqnarray} 
  \vec{S} 
&=&
  \vec{E} \times \vec{H} 
\\
  \vec{S}
&=&
  \frac{1}{4}
  \left[ 
    \left(
      \mathscr{F}^{-1} \left[ \tilde{\alpha}_r^{-1} \right]
      \ast \vec{G}^{+} 
   \right)
   \cdot 
    \left(
      \mathscr{F}^{-1} \left[ \tilde{\beta}_r^{-1} \right]
      \ast \vec{G}^{+} 
   \right)
\right.
\nonumber
\\
&& 
~~~~
\left.
   -
    \left(
      \mathscr{F}^{-1} \left[ \tilde{\alpha}_r^{-1} \right]
      \ast \vec{G}^{-} 
   \right)
    \cdot 
    \left(
      \mathscr{F}^{-1} \left[ \tilde{\beta}_r^{-1} \right]
      \ast \vec{G}^{-}
   \right)
  \right]  \vec{u}
\label{eqn-flux-G2}
.
~~~~
~~~~
\end{eqnarray}
For dispersionless reference parameters,
 this becomes simply
\begin{eqnarray} 
  \vec{S} 
&=&
  \frac{1}{4\sqrt{\epsilon_r \mu_r}}
  \left[ 
    \vec{G}^{+}
   \cdot 
    \vec{G}^{+} 
   -
    \vec{G}^{-} 
    \cdot 
    \vec{G}^{-}
  \right]  \vec{u}
\label{eqn-flux-G2displess}
.
\end{eqnarray}

Since both the $\vec{G}^{\pm} \cdot \vec{G}^{\pm}$ terms 
 are real and positive, 
 we see that $\vec{G}^{+}$ and $\vec{G}^{-}$ contribute positive
 and negative energy fluxes respectively. 
This leads to the simple interpretation that 
 for particular $\vec{E}$ and $\vec{H}$ fields,
 $\vec{G}^{+}$ corresponds to the energy flux directed forward 
 (along $\vec{u}$), 
 and $\vec{G}^{-}$ to flux directed backward.
The need for this distinction between the direction of the 
 flux due to a $\vec{G}^{\pm}$ field, 
 and its direction of travel has already arisen 
 in \ref{Ss-definitions-G} above.

We can also calculate the energy density of the EM field,
 $ \mathscr{U}(t)
=
  \frac{1}{2}
  \epsilon \ast \vec{E}(t) \cdot \vec{E}(t) 
 +
  \frac{1}{2}
  \mu \ast \vec{H}(t) \cdot \vec{H}(t)$.
For transverse fields and a non-dispersive reference, 
 while still allowing for medium dispersion, 
 the energy density in terms of  $\vec{G}^{\pm }$ is
~
\begin{eqnarray} 
  \mathscr{U}
&=&
  \frac{1}{8}
  \left(
    \left[
      \frac{\epsilon}{\epsilon_r}
     +
      \frac{\mu}{\mu_r}
    \right]
    \ast
    \vec{G}^{+}
   \right)
   \cdot 
      \vec{G}^{+}
 +
  \frac{1}{8}
  \left(
    \left[
      \frac{\epsilon}{\epsilon_r}
     +
      \frac{\mu}{\mu_r}
    \right]
    \ast
    \vec{G}^{-}
   \right)
   \cdot 
      \vec{G}^{-}
\nonumber
\\
&&
~~~~
 +
  \frac{1}{8}
  \left(
    \left[
      \frac{\epsilon}{\epsilon_r}
     -
      \frac{\mu}{\mu_r}
    \right]
    \ast
    \vec{G}^{+}
   \right)
   \cdot 
      \vec{G}^{-}
\nonumber
\\
&&
~~~~ ~~~~
 +
  \frac{1}{8}
  \left(
    \left[
      \frac{\epsilon}{\epsilon_r}
     -
      \frac{\mu}{\mu_r}
    \right]
    \ast
    \vec{G}^{-}
   \right)
   \cdot 
      \vec{G}^{+}
.
\label{eqn-energydensity}
\end{eqnarray}

Notice the cross terms, which appear whenever there is a mismatch between
 the reference and medium parameters. 
These occur because of the interference between the 
 $\vec{G}^{+}$ and $\vec{G}^{-}$ contributions to the field.  

For a dispersive reference, 
 the relevant formulae for $\vec{S}$ and $\mathscr{U}$
 are relatively complicated because of the appearance
 of cross terms and/or convolutions.  
However, 
 this should not produce a significant overhead in numerical simulations
 because the code will be switching between time and frequency domains
 at each step, 
 allowing $\vec{S}$ and $\mathscr{U}$ to be calculated in whatever
 way is most efficient.

\subsection{Co-moving frame}\label{Ss-definitions-frame}

We now consider using a moving reference frame.
This is particularly useful in a space-propagated model 
 where the pulse is held as a function of time,
 since it will stay nearly centered when propagating forwards. 
A simple choice of frame speed might be
 the phase velocity at the centre frequency of the pulse, 
 which minimises the motion of the carrier-like oscillations;
 however, 
 the pulse as a whole will move within the frame 
 because of its different group velocity.
The frame translation for a speed $c_f = 1/ \alpha_f \beta_f$
 is
~
\begin{eqnarray}
  t'
&=& 
  t - \gamma / c_f
\\
  \vec{r}' &=& \vec{r}
,
\end{eqnarray}
~
where $\gamma$ is the distance travelled in the direction of $\vec{u}$. 
Thus 
~
\begin{eqnarray}
  \partial_{t}
&=& 
  \partial_{t'}
,
\\
  \nabla
&=&
  \nabla'
 -
  \frac{\vec{u}}{c_f} \partial_t 
.
\label{eqn-z-frametranslation}
\end{eqnarray}

In  vector calculations, we  need to know how this frame translation
 transforms the curl and divergence operations. 
The divergence is a straightforward consequence of 
 eqn. (\ref{eqn-z-frametranslation}),
 and the curl of an arbitrary vector $\vec{Q}$
 transforms to 
~
\begin{eqnarray}
  \nabla \times \vec{Q}
&=&
  \nabla' \times \vec{Q}
 - 
  \alpha_f \beta_f \vec{u} \times \partial_t Q
.
\label{eqn-vector-frametranslation}
\end{eqnarray}

The ratio of the reference speed 
 (the phase velocity in the reference \lq\lq medium\rq\rq 
 ~ described by $\epsilon_r, \mu_r$) 
 and frame speeds is 
~
\begin{eqnarray}
\xi &=& \alpha_f \beta_f / \alpha_r \beta_r
.
\end{eqnarray}

If we choose to give $\alpha_f$ and $\beta_f$ a frequency dependence,
 we have defined a ``dispersive frame'', 
 where different frequency components travel at different speeds. 
In such a frame,
  any matching dispersive evolution 
 (i.e. where $\alpha_f=\alpha_r$ and $\beta_f=\beta_r$)
 results in no change to the pulse profile.
However, 
 at the end of the simulation,
 we need to transform from the dispersive frame
 back into a normal (non-dispersive) laboratory frame.
Moreover, 
 using a dispersive frame can give rise to numerical stability problems.

\section{First order wave equation}\label{S-firstorder}

We now derive a set of
 first-order differential equations for the forward and backward directed
 fields $\vec{G}^{\pm}$, 
 and use the moving frame set out above in 
 eqns. (\ref{eqn-z-frametranslation}, \ref{eqn-vector-frametranslation}). 
We assume that the medium is continuous, 
 so that $\partial_z \epsilon = \partial_z \mu = 0$, 
 where $\partial_q \equiv d/dq$.  
This does not impose a significant restriction in practice, 
 since a simulation propagated forwards in space 
 can easily handle interfaces between different media.

\subsection{Derivation}\label{Ss-firstorder-derivation}

For a vector derivation of propagation equations
 for $\vec{G}^{\pm}$, 
 we start with the two relevant (source free) Maxwell's equations.  
Writing them in frequency space, 
 with $\tilde{\alpha}^2=\tilde{\epsilon}$ 
 and  $\tilde{\beta}^2=\tilde{\mu}$;
 and taking the cross product of $\vec{u}$ 
 and the $\nabla \times \vec{H}$ equation yields
~
\begin{eqnarray}
  \vec{u} \times \left( \nabla \times \vec{H} \right)
&=&
  - \imath \omega \tilde{\alpha}^2 \vec{u} \times \vec{E}
,
\label{eqn-firstorder-dtE}
\\
  \nabla \times \vec{E}
&=&
   + \imath \omega \tilde{\beta}^2 \vec{H}
.
\label{eqn-firstorder-dtH}
\end{eqnarray}
Multiplying respectively by $\tilde{\beta}_{r}$ and $\tilde{\alpha}_{r}$ 
and taking sums and differences leads to
~
\begin{eqnarray}
  \nabla \times \tilde{\alpha}_r \vec{E}
 ~~
 \pm
 ~~
  \vec{u} \times \left( \nabla \times \tilde{\beta}_r \vec{H} \right)
&=&
 +
  \imath \omega \tilde{\alpha}_r \tilde{\beta}^2 \vec{H}
\nonumber
\\
&&
 ~~
 \mp 
 ~~
  \imath \omega \tilde{\beta}_r \tilde{\alpha}^2 \vec{u} \times \vec{E}
.
~~~~ ~~~~
\end{eqnarray}

Noting the similarities between this and eqn. (\ref{eqn-S-defs-Gvector}), 
we now reorganize using standard vector identities for $\nabla
\times ( \vec{A} \times \vec{B} )$ 
and $\vec{u} \times ( \vec{u} \times \vec{A} )$.  
Finally, we arrive at a curl equation for $\vec{G}^{\pm}$,
namely
~
\begin{eqnarray}
  \nabla \times \vec{G}^{\pm}
&=&
 \mp 
  \imath \omega 
  \left\{
    \tilde{\beta}_r \tilde{\alpha}^2 
    ~
    \vec{u} \times \vec{E}
   \pm
    \tilde{\alpha}_r \tilde{\beta}^2 
    \vec{u} \times \left[ \vec{u} \times \vec{H} \right]
  \right\}
\nonumber 
\\
&&
 ~~~~
 +
  \imath \omega 
    \tilde{\alpha}_r \tilde{\beta}^2 
    \vec{u} {G}^{\circ}
 ~~
 \mp
  \nabla {G}^{\circ}
.
\end{eqnarray}

We now separate the correction components 
 (depending on $\tilde{\alpha}_c$, $\tilde{\beta}_c$) 
 from the reference components
 (depending on $\tilde{\alpha}_r$, $\tilde{\beta}_r$), 
 and substitute expressions containing $\vec{G}^{\pm}, {G}^{\circ}$
 by referring to eqns. (\ref{eqn-S-defs-Evector}) 
 and (\ref{eqn-S-defs-Hvector}).  
We also note that the terms involving $\vec{G}^{\pm}$ decouple from 
 those involving ${G}^{\circ}$.
Hence 
~
\begin{eqnarray}
  \nabla \times \vec{G}^{\pm}
&=&
 \mp 
  \imath \omega 
  \tilde{\alpha}_r \tilde{\beta}_r 
  ~
  \vec{u} \times \vec{G}^{\pm}
\nonumber 
\\
&&
 ~~
 \mp 
  \frac{\imath \omega \tilde{\alpha}_c \tilde{\beta}_r}
       {2}
    \vec{u} \times 
    \left[ \vec{G}^{+} + \vec{G}^{-} \right]
\nonumber 
\\
&&
 ~~~
 -
  \frac{\imath \omega \tilde{\alpha}_r \tilde{\beta}_c}
       {2}
    \vec{u} \times \left[ \vec{G}^{+} - \vec{G}^{-} \right]
,
\label{eqn-firstorder-Gpm}
\\
  \nabla {G}^{\circ}
&=&
 \pm
  \imath \omega 
    \tilde{\alpha}_r \tilde{\beta}_r
    \vec{u} {G}^{\circ}
 \pm
  \imath \omega 
    \tilde{\alpha}_r \tilde{\beta}_c
    \vec{u}
    {G}^{\circ}
\label{eqn-firstorder-Gcirc}
.
\end{eqnarray}

For media whose magnetic behaviour is matched perfectly by the 
 reference parameters (i.e. $\tilde{\beta}_c=0$), 
 this simplifies to
~
\begin{eqnarray}
  \nabla \times \vec{G}^{\pm}
&=&
 \mp 
  \imath \omega 
  \tilde{\alpha}_r \tilde{\beta}_r 
  ~
  \vec{u} \times \vec{G}^{\pm}
 ~
 \mp 
  \frac{\imath \omega \tilde{\alpha}_c \tilde{\beta}_r}
       {2}
    \vec{u} \times 
    \left[ \vec{G}^{+} + \vec{G}^{-} \right]
,
~~~~
\label{eqn-firstorder-Gpm-std}
\\
  \nabla {G}^{\circ}
&=&
 \pm
  \imath \omega 
    \tilde{\alpha}_r \tilde{\beta}_r
    \vec{u}
    {G}^{\circ}
.
\label{eqn-firstorder-Gcirc-std}
\end{eqnarray}
For propagation along the $z$ axis, 
 in the plane polarized ($E_x$, $H_y$) limit, 
 the curl becomes $\partial_z$, 
 and $\vec{G}^{\pm}$ is replaced with $G_x^{\pm}$.
In the transverse field case, 
 eqn. (\ref{eqn-firstorder-Gcirc-std}) 
 (or (\ref{eqn-firstorder-Gcirc})) can be ignored.

The equations above are written to suggest a spatially directed propagation 
 (along $\vec{u}$), 
 and indeed are most straightforwardly solved that way.
However, a simple rearrangement
 of the terms leads to a $t$-directed propagation model,
 although, in its present form, 
 the time-like evolution of eqn. (\ref{eqn-firstorder-Gpm-std})
 is obscured.  
Whilst $t$-directed propagation has some advantages, 
 it makes the treatment of dispersion and other time-memory
 effects more demanding,
 as discussed by Kolesik and Moloney \cite{Kolesik-M-2004pre} and 
 Tyrrell et.al. \cite{Tyrrell-KN-2005jmo}.

In the time domain, eqn. (\ref{eqn-firstorder-Gpm-std})
 becomes
~
\begin{eqnarray}
  \nabla \times \vec{G}^{\pm}
&=&
 \pm
  \partial_t
  \left[
    \left(
      \alpha_r \ast \beta_r 
    \right)
    \ast
    \left(
      \vec{u} \times \vec{G}^{\pm}
    \right)
  \right]
\nonumber
\\
&&
 ~
 \pm
  \partial_t
  \left[
    \left(
      \frac{\alpha_c \ast \beta_r}
           {2}
    \right)
    \ast
    \left(
      \vec{u} \times 
      \left[ \vec{G}^{+} + \vec{G}^{-} \right]
    \right)
  \right]
.
~~~~ ~~~~
\label{eqn-firstorder-Gpm-approx-time}
\end{eqnarray}

Note that reversing the direction of propagation by changing $\vec{u}$
 to $-\vec{u}$ reverses the roles of $\vec{G}^{+}$ and $\vec{G}^{-}$.  

\subsection{Longitudinal $\vec{E}$}\label{Ss-firstorder-longitudinal}

In the same way that the construction of $\vec{G}^{\pm}$ ignores the 
 contribution from $\vec{H}$ along the propagation vector $\vec{u}$, 
 and forces us to define ${G}^{\circ}$, 
 eqn. (\ref{eqn-firstorder-dtE}) ignores information about the 
 time-evolution of the longitudinal part of $\vec{E}$.
We can rectify this by using 
 $\nabla \cdot ( \vec{u} \times \vec{H} ) 
 = \vec{u} \cdot ( \nabla \times \vec{H} )$,
 to get
~
\begin{eqnarray}
  \tilde{\alpha}_r
  \nabla \cdot 
  \left[ 
    \vec{G}^{+}   -   \vec{G}^{-}
  \right]
&=&
 - 
  \imath \omega
  \tilde{\alpha}^2 
  \tilde{\beta}_r
  \vec{u} 
  \cdot
  \left[ 
    \vec{G}^{+}   +   \vec{G}^{-}
  \right]
.
~~~~ 
\label{eqn-firstorder-longitudinal}
\end{eqnarray}
This is the difference of the source-free divergences for $\vec{G}^{\pm}$
calculated in eqn. (\ref{eqn-S-defs-divergenceGpm}).
Thus eqns. (\ref{eqn-firstorder-Gpm}, 
 \ref{eqn-firstorder-Gcirc}, \ref{eqn-firstorder-longitudinal})
 provide another way of solving the complete set of
 source-free Maxwell's equations using an alternative basis.
Clearly, however, 
 our basis of $\vec{G}^{\pm}, {G}^{\circ}$ is most useful for 
 fields propagating mainly along one axis (i.e. $\vec{u}$), 
 particularly in the limit of transverse fields, 
 where only eqn. (\ref{eqn-firstorder-Gpm}) needs to be solved.

\subsection{Co-moving frame}\label{Ss-firstorder-moving}

We can transform eqn. (\ref{eqn-firstorder-Gpm}) 
directly into a moving 
frame using eqn. (\ref{eqn-vector-frametranslation}),
which gives
~
\begin{eqnarray}
  \nabla' \times \vec{G}^{\pm}
&=&
 \mp 
  \imath \omega 
  \alpha_r \beta_r 
  \left( 1 \mp \xi \right)
  ~
  \vec{u} \times \vec{G}^{\pm}
\nonumber 
\\
&&
~~~~
 \mp 
  \frac{\imath \omega \alpha_c \beta_r}
       {2}
    \vec{u} \times 
    \left[ \vec{G}^{+} + \vec{G}^{-} \right]
 ~~
\nonumber
\\
&&
~~~~~~
  -
  \frac{\imath \omega \alpha_r \beta_c}
       {2}
    \vec{u} \times \left[ \vec{G}^{+} - \vec{G}^{-} \right]
 ~~
.
\label{eqn-firstorder-Gpm-comoving}
\end{eqnarray}
Here we have not shown the transformed (non-transverse) 
 eqns. (\ref{eqn-firstorder-Gcirc}, \ref{eqn-firstorder-longitudinal})
 in the interest of brevity, 
 but they are easy to calculate if needed.

One nice property of this equation is that matching the 
 frame velocity to the phase velocity causes the carrier-like oscillations 
 in the forward travelling fields ${G}^{+}$ to freeze in place, 
 leaving only the evolution due to the correction terms. 
If we are prepared to make the common assumption of 
 only forward-travelling pulses,
 we will have managed to greatly reduce the rate of change of the fields.
This in turn will allow coarser 
 numerical resolutions to be employed in numerical simulations, 
 leading to significant speed advantages over and above those obtained
 by assuming ${G}^{-}=0$.

\subsection{Time vs space propagation}\label{Ss-firstorder-remarks}

In FDTD solutions of Maxwell's equations, 
 optical pulses travel either forwards or backwards in space 
 as they propagate (or march) forward with time.
However, 
 most nonlinear optical simulations are done in a space-propagated picture; 
 with the consequence that optical pulses travel either forwards 
 or backwards in \emph{ time} 
 as the calculations propagate (march) through space. 

Since we follow the space-propagated picture, 
 the pulse travelling forward in time will be described by ${G}^{+}$, 
 and the one travelling backward by ${G}^{-}$.

Note that any backward travelling pulse in a $z$-propagated picture 
 is travelling backwards in 
 time while propagating forwards in space.
Although at first this might seem non-causal, it is in fact the way
 that the simulation represents a pulse which we would normally 
 describe as propagating backwards (i.e. in the direction $-\vec{u}$).  
This is clear from the wave equations; 
 swapping the sign of the propagation direction $\vec{u}$ 
 swaps the behaviour of $\vec{G}^{+}$ and  $\vec{G}^{-}$.

\subsection{Decoupled Wave Equations}\label{Ss-firstorder-decoupled}

We can make the most of our approach by decoupling $\vec{G}^{+}$
 from $\vec{G}^{-}$, 
 enabling the two first-order coupled Maxwell's eqns.
 (\ref{eqn-firstorder-dtE}, \ref{eqn-firstorder-dtH})
 or $\vec{G}^{\pm}$ eqns. (\ref{eqn-firstorder-Gpm})
 (or co-moving form eqn. (\ref{eqn-firstorder-Gpm-comoving}))
 to be reduced to two uncoupled first-order equations.
The equation describing propagation in the \lq\lq uninteresting \rq\rq
 direction can then be discarded,
 leaving one first-order equation where there were originally two.

This step requires an approximation, 
 although since we can perfectly match the reference parameters 
 ($\alpha_r$, $\beta_r$) to the material dispersion, 
 it is not a very stringent one.
Since the correction parameters $\alpha_c$, $\beta_c$ depend only
 on nonlinear effects, 
 they will in general be small,
 keeping cross coupling between $\vec{G}^{+}$ and $\vec{G}^{-}$ minimal.
Further, 
 whilst the $\vec{G}^{+}$ field will rotate forwards 
 according to its wavevector 
 (i.e. $e^{+\imath k z}$, with $k = \alpha_r \beta_r \omega$), 
 the $\vec{G}^{-}$ field will rotate backwards at the same rate 
 (i.e. at $e^{-\imath k z}$).
This means the correction terms for $\vec{G}^{+}$ will contain both an 
 in-sync component from $\vec{G}^{+}$, 
 and a component from $\vec{G}^{-}$ with a large detuning.
Since this detuning (amounting to $e^{-2 \imath k z})$ will usually be large 
 compared to the spatial bandwidth of the pulse, 
 we can apply a rotating wave approximation and 
 average the $\vec{G}^{-}$ contribution to zero.
After applying the same steps to the $\vec{G}^{-}$ equation as well,
 eqn. (\ref{eqn-firstorder-Gpm}) becomes
~
\begin{eqnarray}
  \nabla \times \vec{G}^{\pm}
&=&
 \mp 
  \imath \omega 
  \tilde{\alpha}_r \tilde{\beta}_r 
  ~
  \vec{u} \times \vec{G}^{\pm}
 ~
\nonumber
\\
&&
 ~~~~
 \mp 
  \frac{\imath \omega \tilde{\alpha}_c \tilde{\beta}_r}
       {2}
    \vec{u} \times 
    \vec{G}^{\pm} 
 \mp
  \frac{\imath \omega \tilde{\alpha}_r \tilde{\beta}_c}
       {2}
    \vec{u} \times \vec{G}^{\pm}
,
~~~~
\label{eqn-firstorder-Gpm-uncoupled}
\end{eqnarray}

\section{Simulating ${G}^{\pm}$}\label{S-simulations}

We now examine the procedure needed to simulate 
 wave propagation using the ${G}^{\pm}$ variables.
This will clarify various practical issues as well as 
 illuminate some of the less-obvious features of our approach. 
We consider a plane-polarized EM wave propagating along $z$
 in a non-magnetic medium with dispersion 
 and a weak nonlinearity. 
Since our aim is to explain the fundamental principles of 
 the use of ${G}^{\pm}$ variables, 
 we first present a number of simple examples.

Our numerical simulations of the G$^{\pm}$ wave equations
 are implemented by straightforward adaption of the 
 PSSD technique \cite{Tyrrell-KN-2005jmo}.
In PSSD, 
 fields are stored as functions of time, 
 and fast Fourier transforms (FFTs) are used to convert to the frequency domain
 for the calculation of pseudospectral derivatives 
 and the effects of dispersion.
This technique allows the simple application of arbitrary dispersion, 
 which becomes a simple multiplication in frequency space.
Fields are then transformed back to the time domain, 
 where the nonlinear effects are calculated, 
 before propagating the fields forward in space. 
Computational details, such as how to design the mesh and control the 
  accuracy of the simulations are well known (e.g. the Courant and 
  Nyquist criteria), 
 and can be found in a range of sources 
 (e.g.  \cite{Fornberg-PGPM,Taflove-Hagness-CE}).

When applied to G$^{\pm}$ fields, 
 the basic spatially-propagated PSSD algorithm does not change, 
 but the wave equation to be solved is now 
 eqn. (\ref{eqn-firstorder-Gpm-std}) instead 
 of Maxwell's equations.
This means that the full flexibility of PSSD 
 is harnessed with the advantages of G$^{\pm}$ fields
 to give a powerful and efficient combination.

\subsection{Simulation speed}\label{Ss-simulations-numerics}

The computational speed of any PSSD-type propagation depends
 primarily on the time spent doing FTs.  
In the PSSD technique described in \cite{Tyrrell-KN-2005jmo}, 
 five FTs are used,
 two forward and back pairs, 
 and one (forward only) to calculate the derivative of the
 electric displacement ${D}$.
If magnetic dispersion were present,
 PSSD would require an extra FT for the magnetic induction ${B}$,
 making six FTs in all.

In contrast, 
 a ${G}^{+}$ simulation requires only three FTs.  
This comprises two forward FTs which are used to calculate 
 the derivative for the dispersion and nonlinearity, 
 and one backward FT is used to change back into the time domain;
 the two derivatives are combined in the frequency domain where the
 problem becomes linear.   
Such a simulation will therefore run 67\% faster than the corresponding
 $E$ and $H$ PSSD algorithm; 
 or 100\% if there is also magnetic dispersion.

To include both ${G}^{+}$ and ${G}^{-}$ would require six FTs; 
 one more than the usual PSSD case, 
 but the same if magnetic dispersion needs to be included.

\subsection{Implementation}\label{Ss-simulations-reference}

As a first step, we divide the total permittivity into three:
 a reference component with constant permittivity $\tilde{\epsilon}_{r}$,
 a linear dispersion correction $\tilde{\epsilon}_c^{D}(\omega)$, 
 and an instantaneous nonlinearity $\tilde{\epsilon}_c^{NL}$; 
 the permeability has the vacuum value $\mu_{0}$. 
The medium properties can therefore be represented in the following fashion
~
\begin{eqnarray}
  \tilde{\epsilon} (\omega)
&=&
  \tilde{\epsilon}_r
 +
  \tilde{\epsilon}_c^D (\omega)
 +
  \tilde{\epsilon}_c^{NL}
\\
&=&
  \tilde{\alpha}_r^2
 +
  \tilde{\alpha}_r \tilde{\alpha}_c^D  (\omega)
 +
  \tilde{\alpha}_r \tilde{\alpha}_c^{NL}
\end{eqnarray}
~
This particular breakdown of $\tilde{\epsilon}$ 
 is for illustrative purposes only; 
 in practice, 
 we would choose a dispersive reference and try to leave
 only nonlinear terms in the correction parameters 
 (i.e. use $\tilde{\epsilon} (\omega)
   = \tilde{\epsilon}_r(\omega) + \tilde{\epsilon}_c^{NL}$).
We might also regroup various terms to optimise the numerical performance.

The first order evolution equation, 
specialized from eqn. (\ref{eqn-firstorder-Gpm-std}),
is
~
\begin{eqnarray}
  \partial_z G_x^{\pm}
&=&
 \mp 
  \imath \omega 
  \tilde{\alpha}_r \tilde{\beta}_r 
  \left( 1 \mp \xi \right)
  ~
  G_x^{\pm}
 ~~
 \mp 
  \frac{\imath \omega \tilde{\alpha}_c^D \tilde{\beta}_r}
       {2}
  \left[ G_x^{+} + G_x^{-} \right]
\nonumber
\\
&&
 ~~
 \mp 
  \frac{\imath \omega \tilde{\alpha}_c^{NL} \tilde{\beta}_r}
       {2}
  \left[ G_x^{+} + G_x^{-} \right]
.
\label{eqn-firstorder-example1}
\end{eqnarray}
where the RHS contains respectively 
 a reference carrier term ($\propto \tilde{\alpha}_r$), 
 a linear dispersion term ($\propto \tilde{\alpha}_c^D$), 
 and a nonlinear polarization term ($\propto \tilde{\alpha}_c^{NL}$).
We integrate forward in $z$ using a split-step method, 
 where each term is integrated through $\delta z$ in sequence.
This procedure is accurate to first order,  
 so we need to ensure $\delta z$ is sufficiently small. 

In this simple case, 
 the reference term merely applies a
 complex rotation to the field in frequency space represented by
~
\begin{eqnarray}
  G_{x1}^{\pm}
&=&
  G_x^{\pm} (z) 
 \times
  \exp
    \left[
     \mp 
       \imath \omega 
       \tilde{\alpha}_r \tilde{\beta}_r 
       \left( 1 \mp \xi \right)
      .
      \delta z
    \right]
\label{eqn-firstorder-example1-G1}
.
\end{eqnarray}

If the frame velocity is chosen to be the same as the
 phase velocity given by the reference parameters, 
 the $G_{x}^{+}$ field no longer undergoes any reference evolution 
 as it propagates.
This contrasts with the usual approach, 
 which is to match the frame velocity to the group velocity.
However, with our choice of reference parameters,
 the group velocity corrections 
 appear in the second RHS term as part of $\tilde{\alpha}_c^D$.
If we were to propagate $G_{x}^{\pm}$ in a group velocity frame, 
 we would retain part of the reference term,
 which would then cancel with part of the group velocity contribution 
 from the dispersion term.
This could lead to a better overall cancellation, 
 just as in the usual $E$ field approaches.  
If that were our aim, 
 we could indeed easily rearrange eqn. (\ref{eqn-firstorder-example1}) 
 to incorporate such a cancellation,
 and then solve the equation appropriately.

The next step is to solve for the linear dispersion 
 $\alpha_c^D= \tilde{\epsilon}_c^D / \alpha_r$.  
Fortunately, 
 this part of the equation is also easy to solve exactly 
 in the frequency domain, 
 through the operation
~
\begin{eqnarray}
  G_{x2}^{\pm}
&=&
  G_{x1}^{\pm}
 \times
  \exp
    \left[
     \mp 
       \imath k
     ~
       G_{x1}^{\pm}
       .
       \delta z
\right.
\nonumber
\\
&&
\left.
     ~
      \mp 
      \frac{\imath \omega \tilde{\epsilon}_c^D(\omega)}
           {2}
      \sqrt{\frac{\mu_r}{\epsilon_r}}
      \left[ G_{x1}^{+} + G_{x1}^{-} \right]
      .
      \delta z
    \right]
.
\label{eqn-firstorder-example1-G2}
\end{eqnarray}

Although both reference and dispersion steps can be solved using exponentials,
 there is an important difference.  
The reference evolution of ${G}^{+}$ depends only on ${G}^{+}$, 
 whereas the dispersion evolution depends on the sum ${G}^{+}+{G}^{-}$, 
 since the dispersion acts on the electric field.  
In a forward-only approximation where ${G}^{-}=0$, 
 it is trivial to combine these first two steps, 
 as in most approaches to solving for the propagation of optical pulses.

The third and final step is performed by transforming into the time domain 
 and solving for the $n$-th order nonlinear effects.  
Since the reference $\alpha_r, \beta_r$ are constants, 
 $\alpha_c^{NL} = \chi^{(n)} E^{n-1} / \alpha_r$, 
 a simple Euler method gives
~
\begin{eqnarray}
  G_{x}^{\pm}(z+\delta z)
&=&
  G_{x2}^{\pm}
 ~~
 \pm
  \frac{\chi^{(n)}}
       {2^n}
      \sqrt{\frac{\mu_r}{\epsilon_r^{n-1}}}
\nonumber
\\
&&
  \times 
  \partial_t
  \left[ G_{x2}^{+} + G_{x2}^{-} \right]^n
  \delta z
. 
\label{eqn-firstorder-example1-G3}
\end{eqnarray}

For a narrow-band field centred at $\omega_0$, 
 the time derivative would be dominated by 
 (and proportional to) $\omega_0$.
In most envelope theories we see only this factor $\omega_0$ 
 in the analogous expression; although correction terms 
 exist for wider-band fields 
 \cite{Brabec-K-1997prl,Kinsler-N-2003pra}.

\subsection{Initial conditions: matching a pulse to the medium}
\label{Ss-firstorder-initial}

Most descriptions of pulse propagation start with initial conditions
 chosen to represent a pulse travelling forward in the medium.
Here we consider how to choose the best initial conditions
 for ${G}^{\pm}$ in the case of a pulse
 travelling \emph{ only} in the forward $\vec{u}$ direction.
They are based on the best practical parameterization of the medium 
 $\tilde{\epsilon}_i(\omega)$, $\tilde{\mu}_i(\omega)$,
 which need not be the same as 
 $\tilde{\epsilon}_r$ and $\tilde{\mu}_r$.
Assuming only the electric field $E(\omega)$ of the pulse is known, 
 the procedure is:

\noindent
(1) Choose $\tilde{\epsilon}_i$ and $\tilde{\mu}_i$ 
 to be as close as possible to 
 the actual medium parameters $\tilde{\epsilon}, \tilde{\mu}$.
One might even try to put the nonlinear properties 
 into $\tilde{\epsilon}_i$ and $\tilde{\mu}_i$ as well, 
 but only if one can get a solution for steps (2) and (3) below
 with this added complication.

\noindent
(2) Calculate $H(\omega)$ corresponding to $E(\omega)$ for a 
 forward travelling pulse, so that a ${G}^{-}$ based on 
 $\tilde{\epsilon}_i, \tilde{\mu}_i$ would be zero:
~
\begin{eqnarray}
  H(\omega)
&=&
  - \sqrt{\frac{\tilde{\epsilon}_i(\omega)}{\tilde{\mu}_i(\omega)}} E(\omega)
.
\label{eqn-initialconds-H}
\end{eqnarray}

\noindent
(3) Calculate an initial ${G}^{\pm}$ using the chosen reference parameters
 $\epsilon_r(\omega)$ and $\mu_r(\omega)$, 
 given our initial $E(\omega)$ and $H(\omega)$ fields:
~
\begin{eqnarray}
  {G}^{\pm}
&=&
  \left[
    \sqrt{\tilde{\epsilon}_r(\omega)} 
   \pm 
    \sqrt{\tilde{\mu}_r(\omega)} 
    \sqrt{\frac{\tilde{\epsilon}_i(\omega)}{\tilde{\mu}_i(\omega)}} 
  \right]
  E(\omega)
.
\label{eqn-initialconds-G}
\end{eqnarray}
Note that step (3) is unnecessary if 
$\tilde{\epsilon}_i=\tilde{\epsilon}_r$ and $\tilde{\mu}_i=\tilde{\mu}_r$.

Notwithstanding step (2), ${G}^{-}$ is only eliminated 
 from the simulation if both $\tilde{\epsilon}_i$ and $\tilde{\mu}_i$ are
 perfect matches to the material parameters.  
If the ($\tilde{\epsilon}_i$, $\tilde{\mu}_i$) values are good, 
 but ($\tilde{\epsilon}_r$, $\tilde{\mu}_r$) less so, 
 a weak ${G}^{-}$ field will co-propagate \emph{ forwards} with ${G}^{+}$, 
 even though the Poynting vector of the ${G}^{-}$ is directed backwards.
If ($\tilde{\epsilon}_i$, $\tilde{\mu}_i$) is a bad match as well, 
 the initial ${G}^{-}$ will have a component that \emph{ travels} backwards, 
 its magnitude corresponding to that of the reflection between 
 a medium with parameters ($\tilde{\epsilon}_i$, $\tilde{\mu}_i$)
 and one with the actual parameters.
Since it is usually possible to include all the linear dispersive properties 
 in $\tilde{\epsilon}_i$, $\tilde{\mu}_i$, 
 any discrepancy is likely to be due to the nonlinear contribution, 
 and consequently very small.

\begin{figure}[ht]
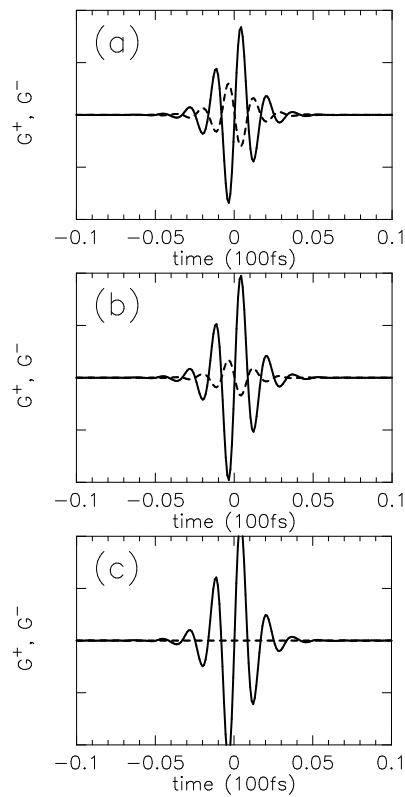

\begin{center}
 \includegraphics[width=0.4\columnwidth,angle=-90]{fig1a-vac.ps}\\
 \includegraphics[width=0.4\columnwidth,angle=-90]{fig1b-nnn.ps}\\
 \includegraphics[width=0.4\columnwidth,angle=-90]{fig1c-perf.ps}
\end{center}
\caption{
A 500nm pulse in fused silica, represented with
(a) a vacuum reference,
(b) a fixed refractive index reference
(c) a perfectly matched dispersive reference.
In all cases the initialization parameters are those that perfectly
match the dispersive properties of the medium.
Solid line: ${G}^{+}$ field. Dashed line: ${G}^{-}$ field.
}
\label{F-diagram-initialconditions}
\end{figure}

Figure \ref{F-diagram-initialconditions} shows how different choices of
 reference parameter affect the ${G}^{\pm}$ fields required to model 
 a simple forward-propagating few-cycle pulse at 500nm in fused silica.
Note that although the ${G}^{\pm}$ fields in 
 (a) and (b) are directly proportional to $E$ and $H$, 
 in (c), the use of a dispersive reference means that a
 deconvolution would be needed (if in the time-domain)
 to transform from ${G}^{\pm}$ to $E$ and $H$.

In figure \ref{F-diagram-initialconditions}(a), 
 the mismatch between the reference parameters (with $n=1$) 
 and the actual medium ($n\approx 1.5$ at the 500 nm pulse center wavelength) 
 causes a significant co-propagating ${G}^{-}$ component to appear; 
 this is improved in (b) where the reference parameters
 specify a constant refractive index close to that at
 the centre frequency of the initial pulse. 
Since the mismatch 
 between the reference and the true material properties
 is due only to the material dispersion, 
 the initial co-propagating ${G}^{-}$ component is smaller in 
 figure \ref{F-diagram-initialconditions}(b) than in 
 figure \ref{F-diagram-initialconditions}(a).
The reduction in the size of ${G}^{-}$ is rather smaller than might
 be expected, 
 mainly because although fused silica has a 
 refractive index of about 1.5 at 500nm, 
 we have used non-dispersive reference parameters with
 a refractive index of 1.5 at all frequencies.

In figure \ref{F-diagram-initialconditions}(b), 
 although the construction of ${G}^{\pm}$ has the phase velocity 
 reasonably well matched, 
 the group velocity of the pulse is poorly matched.  
In addition there is a smaller effect caused by the wide-band nature
 of the pulse, 
 where the reference parameters are (even) less well matched 
 to frequency components away from the center frequency.

The conclusion is that, 
 for any pulse propagating in a material whose dispersion 
 is not perfectly matched by the reference over the pulse bandwidth, 
 a finite co-propagating ${G}^{-}$ will appear. 
This will be made up of frequency components whose phase velocity 
 in the medium do not match the phase velocity given by the reference. 
Thus, in typical dispersive media, 
 only very narrow-band pulses result in a negligible ${G}^{-}$
 for non-dispersive reference parameters. 
However, 
 the mismatch between the reference and the true material properties 
 can be completely removed by using a dispersive reference identical 
 to that of the material being simulated. 
The results of this are shown in 
 figure \ref{F-diagram-initialconditions}(c), 
 where ${G}^{-}$ is  identically zero.

\subsection{Dispersive propagation}\label{Ss-simulations-propagation}

We now present a variety of numerical results demonstrating pulse
 propagation in a dispersive medium.  
We take the medium to have the properties of fused silica, 
 but we do not include nonlinear effects for the moment. 
Our aim is to give a flavour 
 of what the ${G}^{\pm}$ fields look like for different reference parameters.
The choice of reference is important because, 
 as explained earlier, 
 if the reference is not perfectly matched to the actual medium, 
 a forward travelling pulse will contain a ${G}^{-}$ wave 
 co-propagating with the main ${G}^{+}$ component.  
Usually we will want to choose a reference that makes ${G}^{-}$ negligible, 
 so we can save computational effort.

We deliberately choose ultra-short pulses containing only
 a few optical cycles to demonstrate 
 the flexibility of our method in the short pulse limit.

\begin{figure}[ht]
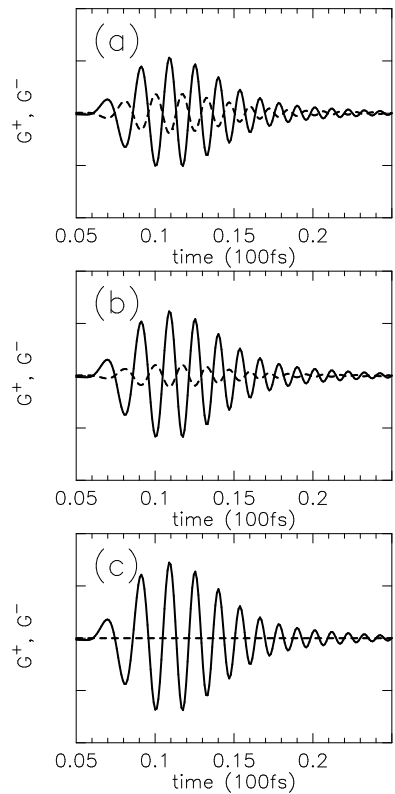

\begin{center}
 \includegraphics[width=0.4\columnwidth,angle=-90]{fig2a-vac.ps}\\
 \includegraphics[width=0.4\columnwidth,angle=-90]{fig2b-nnn.ps}\\
 \includegraphics[width=0.4\columnwidth,angle=-90]{fig2c-perf.ps}
\end{center}
\caption{
The same pulse as in figure \ref{F-diagram-initialconditions}, 
after propagating 15$\mu$m; 
represented in 
(a) a vacuum reference,
(b) a fixed refractive index reference
(c) a perfectly matched dispersive reference.
Solid line: ${G}^{+}$ field. Dashed line: ${G}^{-}$ field.
}
\label{F-diagram-fifteenmicron}
\end{figure}

Figure \ref{F-diagram-fifteenmicron} shows the results for the 
 fields in fig. \ref{F-diagram-initialconditions} after propagating 
 15$\mu$m in fused silica with the nonlinearity ignored.
In all cases, 
 the initial size of ${G}^{-}$ is broadly maintained and, 
 in particular, it remains zero when the reference parameters are 
 perfectly matched.  
Although the ${G}^{\pm}$ fields in  \ref{F-diagram-fifteenmicron}(a) 
 and  \ref{F-diagram-fifteenmicron}(b) are directly proportional 
 to $E$ and $H$,
 in (c), the use of a dispersive reference means that {in that case} 
 a deconvolution is needed
 to transform from ${G}^{\pm}$ to $E$ and $H$.

We can also consider the effect of neglecting a
 finite (but significant) ${G}^{-}$ field,
 where the ${G}^{+}$ part of the pulse then undergoes 
 the wrong dispersion.  
This is because the dispersive correction part 
 (see e.g. eqn. (\ref{eqn-firstorder-example1})) depends on ${G}^{+}+{G}^{-}$, 
 and thus, without ${G}^{-}$, will be either too big or too small.
This problem is avoided by using a dispersive reference identical 
 to that of the material being simulated. 
The results of this are shown in 
 figure \ref{F-diagram-fifteenmicron}(c), 
 where ${G}^{-}$ is always identically zero and 
 no approximation is necessary to omit ${G}^{-}$.

\begin{figure}[ht]
\begin{center}
 \includegraphics[width=0.4\columnwidth,angle=-90]{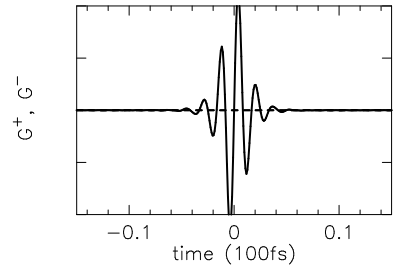}
\end{center}
\vspace{-5mm}
\caption{
Simulation results showing ${G}^{\pm}$ for perfect reference
($\epsilon_r=\epsilon_{silica}(\omega)$) and a
matched dispersive frame.
}
\label{F-diagram-sim-PropPerRef}
\end{figure}

In figure \ref{F-diagram-sim-PropPerRef} we 
 show the result for a simulation with both
 a perfectly matched dispersive reference
 {\em and} a perfectly matched dispersive frame.
Since all the 
 material properties are included in the reference parameters, 
 \emph{ and} we pick a frame that exactly matches the propagation, 
fig. \ref{F-diagram-sim-PropPerRef}
looks identical to the initial state in 
fig. \ref{F-diagram-initialconditions}(c).
We can recover the expected lab-frame final state by transforming 
fig. \ref{F-diagram-sim-PropPerRef} out of its dispersive frame,
and so get a graph identical to 
fig. \ref{F-diagram-fifteenmicron}(c).

The main message from these simulations is that
 the better matched the reference parameters are to the material parameters, 
 the smaller the co-propagating ${G}^{-}$.  
For a perfectly matched reference, 
 the co-propagating ${G}^{-}$ vanishes.
Also, 
 the better matched the frame is to the material parameters, 
 the slower the evolution of the pulse shape.  
However, 
 we then have to do more work to transform the final state of the pulse 
 (in its moving frame)
 into the stationary-frame counterpart we would see in the lab -- 
 although for a linearly dispersive frame, 
 the transformation is straightforward.

\subsection{Nonlinear propagation}\label{Ss-simulations-nonlinearity}

We now demonstrate some simple pulse propagations in nonlinear media.
Since neither the initial conditions (determined by $\epsilon_i, \mu_i$)
 nor the reference parameters (determined by $\epsilon_r, \mu_r$) 
 include the nonlinearity, 
 the pulse is not perfectly forward propagating, 
 and a small ``reflection''
 occurs as the pulse starts propagating in the nonlinear medium.

\begin{figure}[ht]
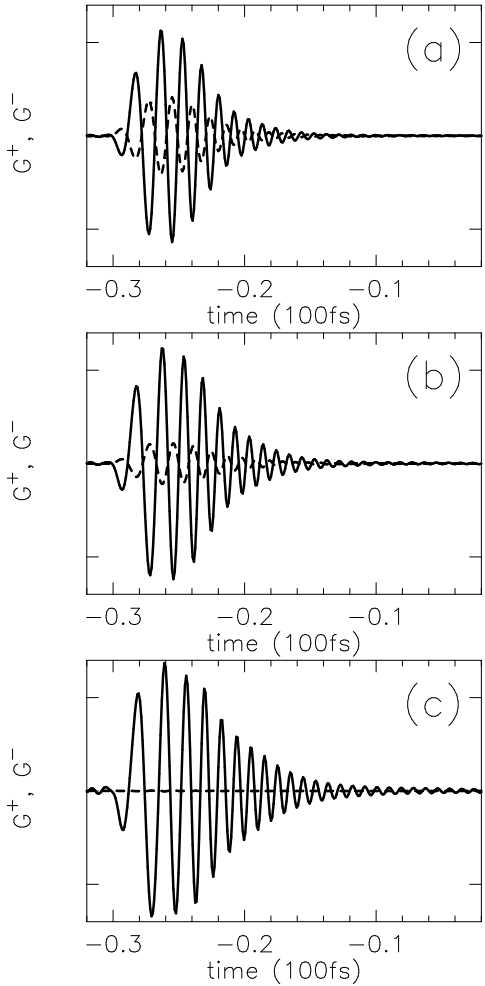

 \includegraphics[width=0.5\columnwidth,angle=-90]{fig4a-chirp-iii.ps}\\
 \includegraphics[width=0.5\columnwidth,angle=-90]{fig4b-chirp--ii.ps}\\
 \includegraphics[width=0.5\columnwidth,angle=-90]{fig4c-chirp--iv.ps}
\caption{
A similar pulse as above, after propagating 10$\mu$m through fused silica; 
represented in 
(a) a vacuum reference,
(b) a fixed refractive index reference ($n=1.5$),
(c) a perfectly matched dispersive reference.
Parameters have been adjusted to give a clearer final pulse shape.
Solid line: ${G}^{+}$ field. Dashed line: ${G}^{-}$ field.
}
\label{F-diagram-sim-nonlinearity-chi3}
\end{figure}

Figure \ref{F-diagram-sim-nonlinearity-chi3} shows
 how pulses similar to those in fig. \ref{F-diagram-initialconditions}
 look after propagating $10\mu$m through fused silica. 
The pulse parameters  were adjusted to give a clearer final pulse shape.
We see the same pattern as in 
 figs \ref{F-diagram-initialconditions} and \ref{F-diagram-fifteenmicron}, 
 where a weak ${G}^{-}$ remains except for 
 perfectly matched reference parameters. 
Note, however, 
 that the addition of nonlinearity does not cause the size of
 the ${G}^{-}$ field to change significantly during propagation.

\begin{figure}[ht]
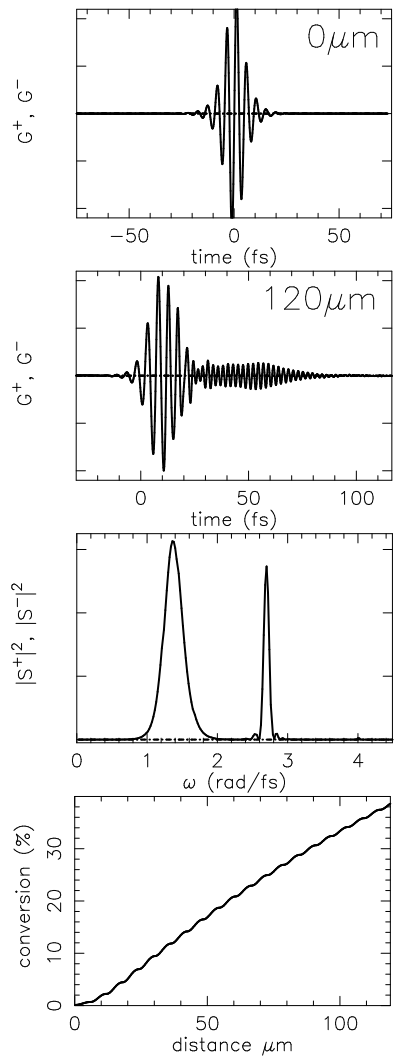

\begin{center}
 \includegraphics[width=0.4\columnwidth,angle=-90]{fig5a-shg2-i.ps}\\
 \includegraphics[width=0.4\columnwidth,angle=-90]{fig5b-shg3-f.ps}\\
 \includegraphics[width=0.4\columnwidth,angle=-90]{fig5c-shg3-fw.ps}\\
 \includegraphics[width=0.4\columnwidth,angle=-90]{fig5d-shg3-cv.ps}
\end{center}
\caption{
Second harmonic generation in 120 $\mu$m of LiNO$_3$, 
periodically poled at 6.05$\mu$m.
Clockwise from top left: 
  initial pulse,
  final pulse,
  second harmonic power,
  final pulse spectrum.
Solid line: ${G}^{+}$ field. Dot-dashed line: ${G}^{-}$ field.
}
\label{F-diagram-nonlinear-shg}
\end{figure}

We now apply our approach to the practical problem of second 
 harmonic generation in 120$\mu$m of periodically poled lithium niobate.
The results are shown on figure \ref{F-diagram-nonlinear-shg}, 
 and agree with the simulations of 
Tyrrell et.al. \cite{Tyrrell-KN-2005jmo}.

\subsection{Some remarks on layered media}\label{Ss-simulations-remarks}

A complication arises when propagating a pulse through
 layers of material with significantly different dispersions. 
Because the $\vec{G}^{\pm}$ definitions are carefully constructed 
 to match the propagation medium, 
 $\vec{G}^{\pm}$ variables ideal for one layer 
 (and so ensuring $\vec{G}^{-}=0$)
 will not be ideal for another.
This gives us two options: 
 (a) either retain the $\vec{G}^{-}$ field in the description, 
 or
 (b) at each layer boundary, switch to a set of $\vec{G}^{\pm}$ variables
 matched to that medium.
Option (a) is simpler, 
 but it is not necessarily computationally efficient 
 and leads to complications involving reflections from the interfaces.
Option (b) is more efficient computationally when we are 
 only interested in the forward-going pulse, 
 as the effort involved in switching $\vec{G}^{\pm}$ definitions is 
 comparable to only a single spatial step 
 in the ongoing propagation calculation.

\subsection{Justifying the forward-only approximation}\label{Ss-simulations-forwardonly}

The ability to accurately incorporate dispersion into our  
 reference permittivity allows great control 
 over the magnitude of the ${G}^{-}$ field.
Our tests have shown that we can confidently neglect ${G}^{-}$
 if our construction of ${G}^{\pm}$ accurately includes 
 the medium dispersion, 
 although possible exceptions may occur 
 in cases involving extremely strong nonlinearities.

This can be seen in the case of periodically poled lithium niobate
 discussed above (see fig. \ref{F-diagram-nonlinear-shg}), 
 where the ratio of the ${G}^{-}$ to ${G}^{+}$ intensities 
 was $1:10^6$.
An even more rigorous test of ${G}^{+}$'s ability to accurately simulate 
 short pulse propagation was our recent study of the effects of 
 dispersion on carrier shocking \cite{Kinsler-TRN-2005-draft}.
Despite the strong nonlinear effects, 
 and significant distortion to the pulse profiles, 
 ${G}^{+}$ simulations consistently produced results in 
 agreement with PSSD --
 whilst still only requiring half the computational effort.

The ability to accurately model pulse propagation using {\em only} ${G}^{+}$
 after carefully choosing a reference permittivity
 clearly justifies neglecting ${G}^{-}$, 
 which in turn simplifies numerical simulations.

\section{Envelope propagation equation}\label{S-envelopes}

When computing the interaction of narrow-band fields, 
 it is common to remove chosen carrier frequencies, 
 and to evolve the envelopes rather than the complete EM fields.
In fact, if sufficient care is taken with the approximations, 
 and the system simulated is well behaved, 
 even quite wide-band pulses can be successfully modelled in this way.

We can use an envelope approach with the ${G}^{\pm}$ variables.
However,
 a full model requires four envelopes to describe the ${G}^{\pm}$,
 just as in a complete Maxwell theory where envelopes are needed for
 both the backward and forward travelling $E$ and $H$.
A full expansion of ${G}^{\pm}$ into forward and backward envelopes
 ${G}^{\pm}_f$, ${G}^{\pm}_b$ would be
~
\begin{eqnarray}
  {G}^{\pm} (\omega)
&=&
    \mathscr{G}^{\pm}_f (\omega \mp \omega_0)
    e^{ \pm \imath k z }
   +
    \mathscr{G}^{\pm}_f ~^* (\omega \mp \omega_0)
    e^{ \mp \imath k z }
\nonumber
\\
&&
 +
    \mathscr{G}^{\pm}_b (\omega \mp \omega_0)
    e^{ \pm \imath k_0 z }
   +
    \mathscr{G}^{\pm}_b ~^* (\omega \mp \omega_0)
    e^{ \mp \imath k_0 z }
,
~~~~ ~~~~ 
\end{eqnarray}
where we have suppressed the $z$ argument on the 
 envelope functions for brevity.  
Note that the forward-like ${G}^{-}$ contribution (i.e. $\mathscr{G}^{-}_f$)
 needs a backward-travelling carrier, 
 as otherwise it is not possible to match the reference evolution terms
 for both $\mathscr{G}^{+}_f$ and $\mathscr{G}^{-}_f$.
When inserted into the wave equations,
  this expansion results in a large number of terms,
  even for the relatively simple case of a third-order nonlinearity.
However, 
 we can specialize to the case where only 
 forward-travelling waves are considered, 
 and set $\mathscr{G}^{\pm}_b=0$.
Since the backward-travelling waves are now eliminated,
 we can propagate pulses efficiently in a moving frame.  
This is important, because the backward parts in a moving frame move
 at \emph{ twice} the frame speed.  
In a full (non-envelope) simulation, 
 we need somehow to filter out the backward components, 
 as otherwise the hoped-for numerical gains are lost 
 by the fact that a finer $z$-step is required for accurate integration.

The first order wave equation for the forward-travelling envelopes
 defined above is
~
\begin{eqnarray}
    \partial_{z} 
      \mathscr{G}^{\pm}_f
&=&
 \mp
  \imath 
  \left( 
    \omega \tilde{\alpha}_r \tilde{\beta}_r 
   -
    k_0
  \right)
        \mathscr{G}^{\pm}_f
 \mp
  \frac{\imath \omega a_c \tilde{\beta}_r}
       {2}
  \left\{
          \mathscr{G}^{\pm}_f
     +
      \mathscr{G}^{\mp}_f ~^*
  \right\}
.
~~~~
\end{eqnarray}
Here $a_c+a_c^*=\tilde{\alpha}_c$, 
 which is simple in the case of dispersion but, 
 in the presence of nonlinearity,
 will be the appropriately carrier-matched, 
 positive frequency part of the permittivity correction parameter.

If $\omega_0 = k_0 c_r$, we have 
~
\begin{eqnarray}
    \partial_{z} 
      \mathscr{G}^{\pm}_f
&=&
 \mp
  \imath \tilde{\alpha}_r \tilde{\beta}_r
  \left( 
    \omega  
   -
    \omega_0
  \right)
        \mathscr{G}^{\pm}_f
 \mp
  \frac{\imath \omega a_c \tilde{\beta}_r}
       {2}
  \left\{
          \mathscr{G}^{\pm}_f
     +
      \mathscr{G}^{\mp}_f ~^*
  \right\}
.
~~~~
\end{eqnarray}
~~~~ ~~~~ ~~~~ ~~~~ ~~~~

In a suitable narrow-band limit, 
 we should be able to ignore the first term on the RHS of this equation,
 leaving the evolution of the envelopes to be controlled solely by
 the correction term.  
The description can be easily generalized to cases involving 
 multiple components centred on different carrier frequencies.
Note that this is a first-order envelope equation, 
 and, as such, 
 does not require the various extra approximations needed
 when deriving an envelope propagation equation from the standard 
 ($E$ field) 
 second order wave equation.

\section{Second order wave equation}\label{S-secondorder}

In section \ref{S-firstorder}
 we derived first-order wave
 equations for the field variables $\vec{G}^{\pm}$.  
However, 
 since many pulse propagation theories start from a second-order form, 
 we have also derived a second-order 
 propagation equation.  
We apply the usual restriction to transverse-only fields, 
 and split the medium properties (i.e. the permittivity and permeability) 
 into a reference part (with $c_r = 1/\alpha_r \beta_r$),
 a linear dispersive part (controlled by $\alpha_c^{D}, \beta_c^{D}$)
 and a nonlinear electro-optic polarization part 
 ($\vec{P} = \alpha_r \alpha_c^{NL}  \ast \vec{E}$).
The time-domain wave equation for a non-dispersive reference is
~
\begin{eqnarray}
  \nabla^2 \vec{G}^{\pm}
 -
  \frac{1}{c_r^2}
  \partial_t^2
  \vec{G}^{\pm}
 ~~~~ ~~~~ ~~~~ ~~~~
 ~~~~ ~~~~ ~~~~ ~~~~
 ~~~~ ~~~~ ~~~~ ~~~~
 && 
\nonumber
\\
 -
  \frac{1}{2}
  \partial_t
  \left\{
    \frac{1}{c_r}
    \partial_t
   \mp
    \vec{u} \times 
    \nabla \times 
  \right\}
 ~~~~ ~~~~ ~~~~ ~~~~
 ~~~~ ~~~~ ~~~~ ~~~~
&& 
\nonumber
\\
 .
  \left\{
    \alpha_c^{D} \ast 
    \left[
      \vec{G}^{+} + \vec{G}^{-}
    \right]
   \pm
    \beta_c^{D} \ast 
    \left[
      \vec{G}^{+} - \vec{G}^{-}
    \right]
  \right\} 
&&
\nonumber
\\
=
 +
  \frac{1}{2\alpha_r}
    \partial_t
  \left[
    \frac{1}{c_r}
    \partial_t
   \mp
    \vec{u} \times 
    \nabla \times
  \right]
      \vec{P}
.
&&
\label{eqn-secondorder}
\end{eqnarray}

This is similar to the usual second-order equation for the electric field,
 but has the addition of a curl operator
 applied to the dispersion and polarization terms.
This second order wave equation can be solved with the use of an
 envelope-carrier representation for $\vec{G}^{\pm}$, 
 as is often done with the standard equation for the electric field $E$.
Such a derivation can be found in \cite{Kinsler-FCPP}, 
 which contains both SVEA and GFEA \cite{Kinsler-N-2003pra} 
 versions for both $E$ and ${G}^{\pm}$.
The most general form of eqn. (\ref{eqn-secondorder}) appears in 
 \cite{Kinsler-FLECK}.

\section{Alternative Definitions}\label{S-alternatives}

Just as one may decide to propagate the $D$ field instead of the $E$ field, 
 so directional field variables in the style of $\vec{G}^{\pm}$ 
 can be defined in a number of ways.
Continuing with the pattern of combining 
 transverse field components with a cross product, 
 alternative directional fields are
~
\begin{eqnarray}
  \vec{G}'^{\pm} 
&=&
  \vec{u} \times \tilde{\alpha}_r \vec{E} + \tilde{\beta}_r  \vec{H}
,
~~~~ ~~~~ ~~
  {G'}^\circ
=
  \vec{u} \cdot \tilde{\alpha}_r \vec{E}
;
\label{eqn-alternative-Gp}
\\
  \vec{F}^{\pm} 
&=&
  \tilde{\alpha}_r^{-1} \vec{D} + \vec{u} \times \tilde{\beta}_r^{-1}  \vec{B}
,
~~~~ ~~
  {F}^\circ
=
  \vec{u} \cdot \tilde{\beta}_r^{-1} \vec{B}
;
\label{eqn-alternative-F}
\\
  \vec{F}'^{\pm} 
&=&
  \vec{u} \times \tilde{\alpha}_r^{-1} \vec{D} + \tilde{\beta}_r^{-1} \vec{B}
,
~~~~ ~~
  {F'}^\circ
=
  \vec{u} \cdot \tilde{\alpha}_r^{-1} \vec{D}
.
\label{eqn-alternative-Fp}
\end{eqnarray}

The $\vec{G}^{\pm}$ or $\vec{G}'^{\pm}$ variables will best suit 
 problems defined in terms of $\vec{E}$ and $\vec{H}$; 
 the $\vec{G}^{\pm}$ are best suited to electric media, 
 and the $\vec{G}'^{\pm}$ to magnetic media.
In contrast,
 the $\vec{F}^{\pm}$ or $\vec{F}'^{\pm}$ variables are more
 suited to $\vec{D}$ and $\vec{B}$.  
All these definitions can be used to generate wave equations, 
 by a similar procedure to that in section \ref{S-firstorder}.
A point to note is that 
 if the wave equations are generalized to include source terms, 
 the $\vec{G}^{\pm}$ and $\vec{G}'^{\pm}$ forms 
 (or $\vec{F}^{\pm}$ and $\vec{F}'^{\pm}$ forms)
 of the wave equations look somewhat different.

As an example, here are the full first order wave equations for the 
 $\vec{F}^{\pm}, {F}^{\circ}$ form, which is conceptually closest
 to the UPPE (unidirectional pulse propagation equation) 
 of Kolesik et.al. \cite{Kolesik-MM-2002prl,Kolesik-M-2004pre}
 based on projections of $D$ --
~
\begin{eqnarray}
  \nabla \times \vec{F}^{\pm}
&=&
 \mp 
  \imath \omega 
  \alpha_r \beta_r 
  ~
  \vec{u} \times \vec{F}^\pm
\nonumber
\\
&&
 ~~
 \mp 
  \frac{\imath \omega \alpha_r \beta_c}
       {2}
    \vec{u} \times 
    \left[ \vec{F}^+ + \vec{F}^- \right]
\nonumber
\\
&&
 ~~
 ~~
 -
  \frac{\imath \omega \alpha_c \beta_r}
       {2}
    \vec{u} \times \left[ \vec{F}^+ - \vec{F}^- \right]
\nonumber
\\
&&
 ~~
 ~~
 ~~
 \pm 
 ~~
  \vec{u} \times 
  \left( \beta_r + \beta_c \right)
  \vec{J}
,
\label{eqn-1storderD-statframe-transverse}
\\
 \pm
  \nabla {F}^{\circ}
&=&
 +
  \imath \omega 
    \alpha_r \beta_r
    \vec{u} ~ 
    {F}^{\circ}
 ~~
 +
  \imath \omega 
    \alpha_c \beta_r
    \vec{u} ~ 
    {F}^{\circ}
,
\label{eqn-1storderD-statframe-longitudinal}
\\
  \nabla 
  \cdot
  \left(
    \vec{F}^{+} - \vec{F}^{-} 
  \right)
&=&
  -
  \imath \omega
  \alpha_r 
  \left( \beta_r + \beta_c \right)
  \vec{u} 
  \cdot
  \left(
    \vec{F}^{+} + \vec{F}^{-} 
  \right)
\nonumber
\\
&&
 ~~
 +
 \left( \beta_r + \beta_c \right)
  \vec{u} 
  \cdot
  \vec{J}
.
\label{eqn-1storderD-statframe-divergence}
\end{eqnarray}

\section{Conclusions}\label{S-conclusion}

We have introduced generalized forms of the
 directional field variables first envisaged by 
 Fleck\cite{Fleck-1970prb}].  
We have demonstrated that they are
 associated with energy fluxes in the forward and backward directions.
They provide the ideal basis for the standard ``forward-only''
 pulse propagation model,
 both improving our insight into pulse propagation, 
 and allowing the backward-propagating component
 to be efficiently discarded if desired.  
By developing the theory in frequency space, 
 we have shown how the dispersive properties of the
 propagation medium can be incorporated.

We have derived first-order wave equations for $\vec{G}^{\pm}$ 
 that are equivalent to Maxwell's equations.
If dispersion is included carefully, 
 the equations decouple
 and we can get a single equation for forward-only propagation, 
 and hence achieve significant speed gains over
 direct Maxwell's equation solvers for $E$ and $H$ 
 (such as PSSD\cite{Tyrrell-KN-2005jmo} or FDTD\cite{Yee-1966iee}).
We have also presented a number of simulations demonstrating their use.  

Since the $\vec{G}^{\pm}$ variables are not restricted to use
 in first order wave equations,
 we have also presented
 an envelope theory and a second-order wave equation 
 analogous to those regularly used in pulse propagation work.
Either of these equations can be used to extend the practical applications 
 of $\vec{G}^{\pm}$ variables
 into the long-pulse narrow-band regimes.
Further, 
 $\vec{G}^{\pm}$ can still be constructed from the $E$ and $H$
 field obtained in traditional simulations, 
 enabling their use for either diagnosis or analysis.

\begin{acknowledgments}
We would like to acknowledge useful discussions with J.C.A. Tyrrell.
\end{acknowledgments}


\end{document}